\documentclass[conference]{IEEEtran}
\IEEEoverridecommandlockouts
\usepackage{cite}
\usepackage{amsmath,amssymb,amsfonts}
\usepackage{algorithmic}
\usepackage{graphicx}
\usepackage{textcomp}
\usepackage{xcolor}
\usepackage{float}
\usepackage{tabularx} 
\usepackage{hyperref}
\usepackage{longtable}
\usepackage{booktabs}
\usepackage{soul}
\usepackage[draft=true]{minted}

\usepackage{adjustbox} 

\def\BibTeX{{\rm B\kern-.05em{\sc i\kern-.025em b}\kern-.08em
    T\kern-.1667em\lower.7ex\hbox{E}\kern-.125emX}}

\begin{document}
\pagestyle{plain}
\title{ChatNVD: Advancing Cybersecurity Vulnerability Assessment with Large Language Models
}

\author{%
    \begin{minipage}{1.0\textwidth}
        \centering
        Shivansh Chopra$^{\ast}$,
        Hussain Ahmad$^{\ast\dagger}$,
        Diksha Goel$^{\ddagger}$,
        Claudia Szabo$^{\ast}$ \\
        
        $^{\ast}$The University of Adelaide, Australia \\
        $^{\ddagger}$CSIRO's Data61, Australia \\
        shivansh.chopra@student.adelaide.edu.au, \{hussain.ahmad, claudia.szabo\}@adelaide.edu.au, diksha.goel@csiro.au
    \end{minipage}
    \thanks{$^{\dagger}$ Corresponding author}
}

\maketitle

\begin{abstract}

The increasing frequency and sophistication of cybersecurity vulnerabilities in software systems underscores the need for more robust and effective vulnerability assessment methods. However, existing approaches often rely on highly technical and abstract frameworks, which hinder understanding and increase the likelihood of exploitation, resulting in severe cyberattacks. In this paper, we introduce ChatNVD, a support tool powered by Large Language Models (LLMs) that leverages the National Vulnerability Database (NVD) to generate accessible, context-rich summaries of software vulnerabilities. We develop three variants of ChatNVD, utilizing three prominent LLMs: GPT-4o Mini by OpenAI, LLaMA 3 by Meta, and Gemini 1.5 Pro by Google. To evaluate their performance, we conduct a comparative evaluation focused on their ability to identify, interpret, and explain software vulnerabilities. Our results demonstrate that GPT-4o Mini outperforms the other models, achieving over 92\% accuracy and the lowest error rates, making it the most reliable option for real-world vulnerability assessment.

\end{abstract}

\begin{IEEEkeywords}
Large Language Model, Cybersecurity, Software Security, Chatbot, Vulnerability Assessment
\end{IEEEkeywords}

\section{Introduction}

Large Language Models (LLMs), a subset of generative artificial intelligence, have revolutionized numerous fields by enabling advanced content generation and process automation \cite{neupane2023impacts, ahmad2025future}. In education, LLMs enhance efficiency in tasks such as generating questions \cite{elkins2023useful}, grading essays \cite{yan2024practical}, providing personalized feedback \cite{guo2024resist}, and evaluating assignments \cite{hsiao2023developing}. Within the entertainment industry, they exhibit remarkable capabilities, crafting video game narratives \cite{latouche2023generating} and generating music captions \cite{deng2023musilingo}, thereby enriching creative workflows. Business operations have also undergone significant transformation through the integration of LLMs, which streamline marketing campaigns \cite{yang2023against}, improve customer service interactions \cite{pandya2023automating}, and optimize supply chain processes \cite{kosasih2024review}. In healthcare, LLMs empower professionals by offering real-time clinical decision support \cite{rao2023evaluating}, advancing medical education \cite{kuckelman2024assessing}, and enabling predictive analytics for disease progression \cite{shoham2023cpllm}. These applications highlight the profound and multifaceted impact of LLMs, positioning them as key drivers of innovation across diverse disciplines \cite{haque2022think}.

The dynamic and ever-evolving nature of the cybersecurity domain demands continuous adaptation to cutting-edge advancements \cite{ullah2025skills}, as each technological innovation introduces potential new attack vectors and expands the surface available for exploitation \cite{da2024survey, goel2024machine}. Software vulnerabilities, whether accidental oversights or deliberate flaws, serve as entry points for system manipulation by threat actors \cite{jayalath2024microservice, thakur2024cyber}. These vulnerabilities can result in severe consequences, including unauthorized data access or alteration, system disruptions through denial-of-service attacks, or the execution of malicious code, potentially enabling full system compromise \cite{ahmad2023review}. To address these risks, organizations frequently conduct proactive evaluations of system resilience through vulnerability assessments \cite{abdulsatar2024towards}. Such assessments play a vital role in fortifying system defences and safeguarding the integrity and security of digital infrastructures \cite{teichmann2023overview}.

\begin{figure}[t]
  \centering
  \includegraphics[width=\columnwidth]{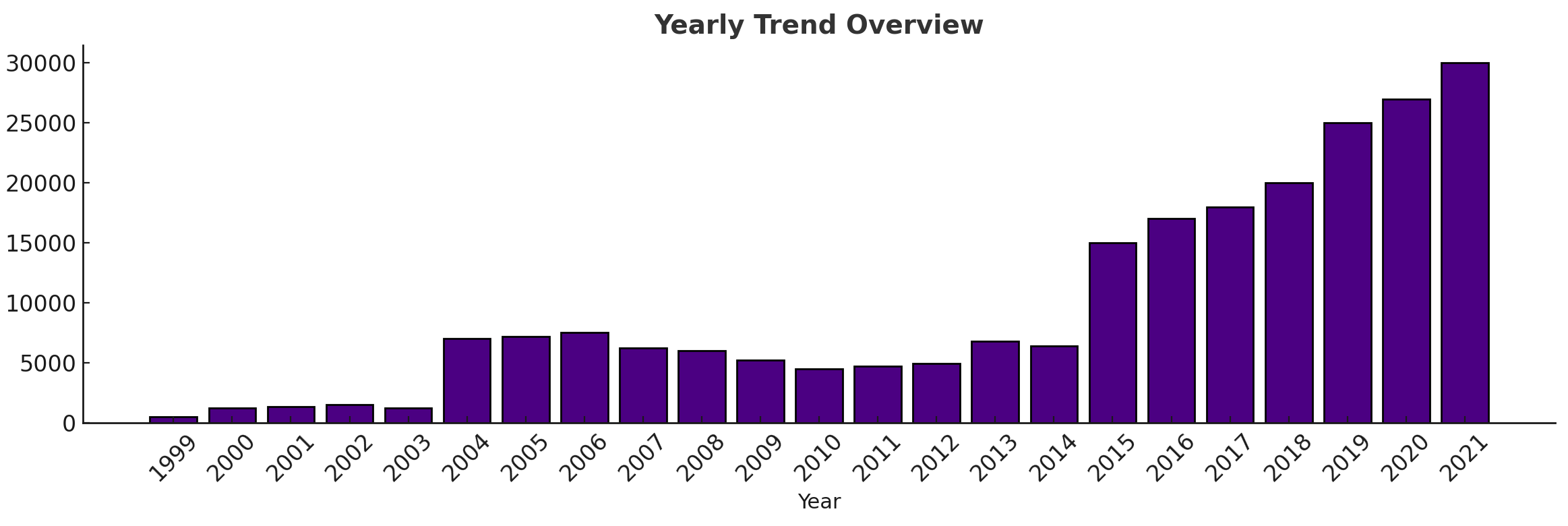}
  \caption{Growth Trend in Reported Software Vulnerabilities \cite{guo2024outside}.}
  \label{fig1}
\end{figure}

The rapid growth in the volume and sophistication of cyber threats has driven a substantial increase in the number of reported software vulnerabilities within the National Vulnerability Database\footnote{\url{https://nvd.nist.gov}}\cite{jimmy2024cyber}. As shown in Figure  \ref{fig1}, the trend of identified vulnerabilities has quadrupled over the past decade \cite{guo2024outside}, reflecting the growing complexity of modern software systems and the expanding attack surface. This surge poses challenges for cybersecurity practitioners in keeping up with evolving threats. According to a recent report by the World Economic Forum, cyberattacks are among the top five global risks in terms of likelihood and impact \cite{mclennan2021global}. Despite this alarming trend, existing methods for vulnerability assessment often depend on overly technical and abstract frameworks, such as manual analysis \cite{austin2013comparison, harer2018automated} and rule-based approaches \cite{afzal2022rule}. These methods are not only time-intensive but also susceptible to human error \cite{harer2018automated}, limiting their effectiveness in addressing the rapidly growing number of vulnerabilities. Furthermore, their complexity and technical nature can create barriers to broader understanding among diverse stakeholders, complicating efforts to prioritize and mitigate vulnerabilities effectively \cite{openai2024gpt4o}. This gap in accessibility and clarity increases the likelihood of exploitation, escalating the risk of severe cyberattacks. To combat this, there is a critical need for more comprehensible and actionable solutions that enable both technical and non-technical stakeholders to assess and mitigate software vulnerabilities effectively.

The NVD serves as a comprehensive repository of software vulnerabilities, offering detailed descriptions that are invaluable for reference \cite{byers2022national}. However, the NVD is designed primarily for lookup and reference purposes and lacks interactive querying capabilities and deeper contextual understanding \cite{NVD-Limitations}. This limitation underscores the need for an advanced solution that goes beyond simple lookup by delivering detailed, context-aware responses tailored to specific vulnerability queries. LLMs can form the backbone of such a tool as they can analyze and interpret NVD data, enabling dynamic interaction with the information and transforming it into a more practical and accessible resource. This approach is especially beneficial for cybersecurity professionals who require a quick, accurate, and in-depth understanding of vulnerabilities to make informed decisions. The NVD, with its extensive dataset of software vulnerabilities, serves as an ideal foundation for training LLMs, enabling the processing of large volumes of data, summarizing complex technical information, and providing actionable insights. An LLM-powered solution would facilitate improved vulnerability comprehension, support more effective mitigation strategies, and reduce the risk of exploitation, benefiting a diverse range of stakeholders, from technical experts to non-technical users.


In this paper, we introduce ChatNVD, an LLM-powered tool designed to enhance software vulnerability assessment by leveraging vulnerability data from the NVD. While LLMs have shown promise in a variety of natural language understanding tasks \cite{vogelsang2025using}, to the best of our knowledge, no prior work has systematically evaluated or benchmarked the applicability of state-of-the-art LLMs to vulnerability data in the NVD for real-world assessment tasks. To address this gap, we develop three variants of ChatNVD using three widely adopted models: GPT-4o Mini (OpenAI) \cite{koubaa2023gpt}, Gemini 1.5 Pro (Google) \cite{GeminiModel_Indepth}, and LLaMA 3 (Meta) \cite{raiaan2024review}. Each model is integrated with a TF-IDF embedding pipeline, selected for its computational efficiency and low-cost processing, as generating high-quality semantic embeddings for the full NVD dataset (720.7 MB) proved prohibitively expensive and time-consuming. We construct a rigorous evaluation framework comprising 125 vulnerability-focused questions derived from real CVEs, covering both factual and contextual queries. Our comparative analysis assesses each model’s ability to retrieve, interpret, and reason over NVD data to provide accurate and actionable insights. This study not only highlights performance disparities across LLMs in vulnerability assessment tasks but also provides concrete evidence of their strengths, limitations, and practical suitability in cybersecurity workflows. Our contributions are twofold:

\begin{itemize}   
    \item We develop ChatNVD, an LLM-powered cybersecurity vulnerability assessment tool built on the dataset of NVD to facilitate a more interactive and efficient approach to provide detailed, context-rich insights and streamline vulnerability analysis for diverse users, including cybersecurity professionals, developers, and non-technical stakeholders.

    \item We perform a comparative analysis of three popular LLMs, GPT-4o Mini, Gemini 1.5 Pro, and LLaMA 3, to assess their capabilities in understanding, processing, and providing accurate and contextually relevant responses to queries about software vulnerabilities. Our evaluation across 125 CVE-based queries shows that GPT-4o Mini achieved the highest overall accuracy (over 92\%), lowest hallucination rate, and most consistent performance across question types and input sizes.
\end{itemize}

The structure of the paper is organized as follows: Section \ref{Section 2} provides a review of related work, covering existing LLMs, embedding techniques, and key cybersecurity insights. Section \ref{Section 3} details the research methodology employed in the development of ChatNVD. Section \ref{Section 4} presents the experimental evaluation, including results and a discussion comparing the three different variants of ChatNVD. Section \ref{implication} outlines practical insights for applying and evaluating LLMs in real-world cybersecurity settings. Section \ref{Section 5} addresses potential threats to validity. Finally, Section \ref{sec:conclusion} concludes the paper and highlights future research directions.


\begin{table*}[h]
\scriptsize
\centering
\caption{\textbf{Comparison of Large Language Models}}
\begin{tabularx}{\textwidth}{l X X X l l c l}
\toprule
\textbf{LLM} & \textbf{Characteristics} & \textbf{Strengths} & \textbf{Limitations} & \textbf{Model Size} & \textbf{References} & \textbf{Open Source} & \textbf{Token Cost (Input/Output)} \\
\midrule
\textbf{GPT-4o Mini} & Fast, scalable, multimodal & High efficiency, low latency, 128K token window & No fine-tuning, limited domain adaptability & Not disclosed & \cite{koubaa2023gpt, yenduri2024gpt, ChatGpt_Strength_1_Step_By_step_Research, ChatGpt_scalable, ChatGpt_Faster_lessResouceIntensive} & No & \$0.15 / \$0.60 per 1M tokens \cite{openai-pricing} \\
\textbf{LLaMA 3} & Lightweight, multilingual & Open-source, efficient, community-supported & Requires domain-specific fine-tuning & 7B parameters & \cite{raiaan2024review, conneau2020unsupervised, brown2020language, Llam_opensource, llama_needs_resources, llama_domain_specific} & Yes & Free \\
\textbf{Gemini 1.5 Pro} & High-performance, multimodal & Accurate, handles text+images, strong context handling & Resource-heavy, limited transparency & 1B+ parameters & \cite{gemini_high_resource, Gemini_multimodal, Gemini_Good_architecture} & No & \$1.25 / \$5.00 per 1M tokens \cite{google-ai-pricing} \\
\bottomrule
\end{tabularx}
\label{llm_comparision}
\end{table*}

\section{Background} \label{Section 2}

This section reviews recent advances in LLMs and text embedding techniques, providing the foundation and motivation for the approach adopted in this study.

\subsection{Large Language Models}

\begin{figure}[!b]
  \centering  \includegraphics[height=1.2\linewidth,width=1.0\columnwidth]{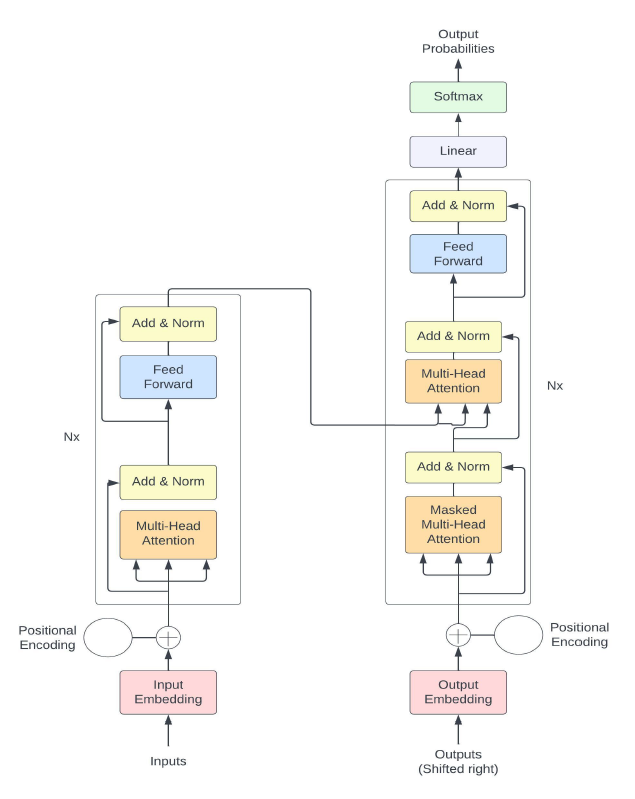}
  \caption{High-Level View of Transformer Architecture Underlying Modern LLMs  \cite{ahmed2023chatgpt}}.
  \label{fig:transformer}
\end{figure}

LLMs represent a major advancement in artificial intelligence, capable of understanding and generating human-like text by learning from vast and diverse linguistic datasets \cite{llm_introduction}. These models are primarily built on the Transformer architecture, which leverages self-attention and multi-head attention mechanisms to effectively capture long-range dependencies and nuanced semantic relationships in language \cite{vaswani2017attention}. Positional encodings further enable Transformers to retain sequence information and preserve semantic integrity, thereby excelling in tasks such as text generation, translation, and summarization \cite{llm_introduction, chang2024survey}. In the domain of cybersecurity, LLMs have been applied in threat intelligence, penetration testing, adaptive intrusion detection, and autonomous defense mechanisms. Their capacity to process real-time data and understand context-rich language makes them particularly effective against sophisticated threats such as advanced persistent threats (APTs). When integrated with established cybersecurity frameworks like MITRE ATT\&CK and NIST, LLMs can support continuous vulnerability scanning \cite{haq2022insider}, risk prioritization \cite{verma2024software}, and intelligent response automation \cite{LLM_Application_Digital_Defense, abbas2023ss}. Figure~\ref{fig:transformer} illustrates the core Transformer architecture that underpins most modern LLMs. While numerous LLMs have emerged in recent years, this study focuses on three leading models developed by major technology companies, GPT-4o Mini (OpenAI), LLaMA 3 (Meta), and Gemini 1.5 Pro (Google), which are described in detail below. A comparative summary of these models is presented in Table~\ref{llm_comparision}.

\vspace{0.07in}

\subsubsection{GPT-4o Mini}

GPT-4o Mini is a cost-efficient variant of the GPT-4 series developed by OpenAI. Using model distillation, it preserves the core capabilities and accuracy of GPT-4 while providing a more favourable performance-cost balance. Its key features include an extensive contextual range and versatile multimodal input support. The model's large 128K-token context window enables efficient processing of long sequences, which is beneficial for tasks such as extended conversation analysis and large-scale dataset interpretation. Furthermore, its ability to process both text and images enhances its utility across diverse applications, from multimedia research to advanced customer support \cite{gpt4mini_datacamp, gpt4mini_official_website}.

\vspace{0.07in}

\textit{Strengths:}
GPT-4o Mini offers economic viability and robust performance. With pricing set at approximately 15 cents per million input tokens and 0 cents per million output tokens, it is more accessible than earlier models like GPT-3.5 Turbo. Empirical evaluations show competitive performance with scores of 82\% on MMLU (textual reasoning), 87\% on MGSM (mathematical reasoning), and 87.2\% on HumanEval (coding proficiency). Additionally, advanced safety mechanisms, such as content filtering, reinforcement learning with human feedback (RLHF), and hierarchical instructions to prevent prompt injection, ensure secure and reliable outputs in public-facing applications \cite{gpt4mini_official_website}.

\vspace{0.07in}

\textit{Weaknesses:}
Despite its advantages, GPT-4o Mini has limitations. Its token-based pricing structure can result in high costs for applications with intensive usage or high traffic. Additionally, although it supports an extensive context window, the maximum output is limited to 1K tokens per request, potentially necessitating multiple API calls for larger outputs. Finally, the lack of fine-tuning capabilities restricts its adaptability for specialized, domain-specific tasks.

\vspace{0.07in}

\begin{table*}[h]
\scriptsize
\centering
\caption{\textbf{Comparison of Text Embedding Techniques}}
\begin{tabularx}{\textwidth}{l l X X c l}
\toprule
\textbf{Technique} & \textbf{Type} & \textbf{Strengths} & \textbf{Limitations} & \textbf{Compute Load} & \textbf{Cost} \\
\midrule
TF-IDF & Statistical, term-based & Fast, interpretable, low-cost & No semantic understanding & Low & Free \\
OpenAI (text-embedding-3-small) & Neural, deep & Captures semantics, few-shot capable & Costly, may overfit & High & \$0.010 / 1M tokens \cite{openai-pricing} \\
LLaMA 3 Embeddings & Neural, multilingual & Adaptable, open-source & Requires tuning; shallow pretraining & High & Free \\
Gemini Embeddings & Multimodal, context-aware & Strong context; handles text + images & High compute, expensive & High & \$0.000025 / 1K chars \cite{google-ai-pricing} \\
\bottomrule
\end{tabularx}
\label{embedding_comp}
\end{table*}

\subsubsection{LLaMA 3}

LLaMA 3 is a large language model developed by Facebook AI Research (FAIR) and built on the RoBERTa architecture, comprising approximately 8 billion parameters \cite{raiaan2024review}. Trained on a diverse corpus, including web pages, books, and Wikipedia articles, LLaMA 3 excels in a wide range of NLP tasks. The model is notable for its state-of-the-art performance, open-source accessibility, and efficient transformer-based design, making it a valuable tool for numerous NLP applications. However, its high resource requirements and potential challenges in specialized domains require careful consideration regarding deployment environments and application contexts.

\vspace{0.07in}

\textit{Strengths:}
LLaMA 3 consistently demonstrates strong performance across multiple NLP benchmarks, excelling in text generation, summarization, and translation \cite{brown2020language}. Its open-source nature promotes community collaboration and innovation, allowing researchers and developers to contribute enhancements and expand its capabilities \cite{Llam_opensource}. Moreover, its optimized transformer-based architecture improves both training speed and inference efficiency, enabling the model to process large datasets in parallel and produce faster, more accurate predictions. Its robust multi-task learning capabilities further allow it to handle a variety of NLP tasks simultaneously, increasing its overall utility.

\vspace{0.07in}

\textit{Weaknesses:}
Despite its efficiency improvements over predecessors, LLaMA 3 remains resource-intensive, requiring substantial computational power for large-scale deployments. This can be challenging for smaller organizations or those with limited access to high-end hardware \cite{llama_needs_resources}. Additionally, while the model is versatile, it may struggle with tasks demanding deep domain-specific knowledge, particularly in specialized fields where high-quality training data is scarce or narrowly focused \cite{llama_domain_specific}.

\vspace{0.07in}

\subsubsection{Gemini 1.5 Pro}

Developed by Google DeepMind, Gemini 1.5 Pro is an advanced iteration within the Gemini language model family, tailored for high-performance NLP tasks \cite{GeminiModel_Indepth}. Its key features include advanced multimodal capabilities, enabling the processing of various data formats, and in-context learning that allows the model to rapidly acquire new skills and adapt to different tasks based solely on prompt exposure, thereby reducing the need for extensive fine-tuning \cite{gemini_official_website}. This combination of multimodal integration and adaptable learning makes Gemini 1.5 Pro a robust solution for cutting-edge NLP applications despite challenges related to resource intensity and interpretability.

\vspace{0.05in}

\textit{Strengths:}
Gemini 1.5 Pro demonstrates high accuracy and superior language understanding, which makes it particularly effective for tasks such as summarization, translation, and complex text generation. Its extensive parameter count enables the model to capture nuanced language differences, yielding outputs that are both contextually appropriate and precise \cite{Gemini_Good_architecture}. Additionally, its ability to seamlessly process and combine diverse data types, including text, images, and voice, broadens its applicability across various domains, from multimedia research to advanced customer support \cite{Gemini_multimodal}.

\vspace{0.07in}

\textit{Weaknesses:}
Despite its advantages, Gemini 1.5 Pro has notable challenges. Its substantial computational requirements for both training and real-time inference may impede adoption, particularly for organizations with limited infrastructure budgets \cite{gemini_high_resource}. Furthermore, the complexity of its internal decision-making processes often results in reduced interpretability, which can be problematic in fields like healthcare and law where understanding the rationale behind model outputs is critical for building trust and ensuring accountability \cite{gemini_lack_transparency}.

\vspace{0.07in}

\subsection{Text Embeddings}

Text embeddings convert words, sentences, or documents into numerical vectors in a multi-dimensional space, capturing semantic relationships and meanings \cite{TextEmbeddings-1}. This allows for the analysis of textual data, while preserving contextual meaning. Recent advancements in embedding techniques, from traditional methods like TF-IDF to transformer-based models, have improved the representation of textual information. Modern models such as BERT, GPT, LLaMA 3, and Gemini use self-attention mechanisms to encode context-dependent information, positioning semantically similar texts closer together and enhancing natural language processing performance \cite{vaswani2017attention}. This section provides an overview of various embedding techniques, with Table \ref{embedding_comp} presenting a comparative analysis of these models.

\vspace{0.07in}

\subsubsection{TF-IDF Embeddings}

Term Frequency-Inverse Document Frequency (TF-IDF) is a numerical statistic that measures the importance of a word within a document relative to its prevalence across a larger corpus. By assigning weights that increase with the word’s frequency in a document and decrease with its commonality in the entire collection, TF-IDF converts textual data into weighted numerical vectors. This process emphasizes unique or significant terms while downplaying ubiquitous ones, making TF-IDF an essential tool in information retrieval and natural language processing \cite{Tf-idf_Definition, Tf-idf_working, ramos2003using}. Its simplicity, computational efficiency, and inherent interpretability allow it to scale to large datasets without demanding extensive computational resources, while higher weights directly indicate more important terms within documents \cite{Tf-idf_interperability, tf-idf_cost_effective}.

\vspace{0.07in}

\textit{Strengths:}
The primary strength of TF-IDF is its efficiency in identifying and quantifying the significance of words. Its straightforward implementation makes it easy to integrate into various NLP pipelines, and its explicit weighting scheme facilitates transparency, which is critical for exploratory data analysis and preliminary feature engineering in machine learning models \cite{tf-idf_book_strength_limitation}.

\vspace{0.07in}

\textit{Weaknesses:}
TF-IDF does not capture semantic meaning or the contextual relationships between words. By treating each term independently, it overlooks syntactic and semantic nuances, reducing its effectiveness in tasks that require deep contextual analysis. In such cases, more sophisticated embedding techniques may be necessary \cite{tf-idf_book_strength_limitation}.

\vspace{0.07in}

\textit{Use Cases:}
TF-IDF embeddings are widely used in text classification, where documents are transformed into numerical features that allow machine learning algorithms to differentiate categories based on term importance. They also play a key role in sentiment analysis by highlighting words that significantly contribute to the conveyed sentiment and in information retrieval systems that rank documents by relevance in response to user queries \cite{Tfidf_classification, Tfidf_sentiment_analysis}.

\vspace{0.07in}

\subsubsection{OpenAI Embeddings}
OpenAI embeddings are sophisticated text representations generated by large language models such as GPT-3. Engineered to capture deep semantic relationships and rich contextual information from extensive corpora, these embeddings support a broad range of NLP tasks, including text generation, summarization, and complex question answering \cite{openAi_embedding}. They also excel in few-shot learning scenarios, adapting to new tasks with minimal additional training data.

\vspace{0.07in}

\textit{Strengths:}  
OpenAI embeddings effectively capture intricate semantic meanings, making them ideal for tasks that demand a nuanced understanding of language. Their versatility and adaptability, along with support for few-shot learning, enable quick adjustment to new tasks without requiring extensive retraining.

\vspace{0.07in}

\textit{Weaknesses:}  
Training and deploying large-scale models like GPT-3 involves substantial computational and financial costs, often necessitating access to specialized infrastructure. These models are pretrained on vast datasets, typically hundreds of billions of tokens, making them impractical to retrain in data-scarce domains. Moreover, while pretraining promotes generalization, improper or narrow fine-tuning can lead to overfitting, potentially reducing performance on unseen or diverse tasks.

\vspace{0.07in}

\textit{Use Cases:}  
OpenAI embeddings are ideally suited for applications demanding comprehensive natural language understanding, ranging from text generation and summarization to complex question answering. Their capacity for few-shot learning makes them particularly effective when domain-specific fine-tuning is minimal.

\vspace{0.07in}

\subsubsection{LLaMA Embeddings}
LLaMA embeddings are generated by a transformer-based language model developed by Facebook AI Research and pre-trained on a massive multilingual corpus. Using a masked language modelling objective, the model learns meaningful representations of words across different languages. These embeddings balance computational efficiency and performance, making them well-suited for large-scale and real-time applications. They are inherently adaptable to multilingual and cross-domain scenarios, and the architecture is optimized for fine-tuning to create task-specific embeddings even with relatively small, domain-specific datasets \cite{pvribavn2024comparative, llama_embedding, dreview}.

\vspace{0.07in}

\textit{Strengths:}
LLaMA embeddings deliver high-quality representations while minimizing computational overhead, which is beneficial in resource-constrained environments. Their multilingual capability enhances versatility for processing diverse languages, and their optimized fine-tuning allows for customization with minimal additional data \cite{llama_embedding}.

\vspace{0.07in}

\textit{Weaknesses:}
Compared to proprietary models like GPT-3, LLaMA models often require more hands-on tuning and domain-specific adaptation to achieve optimal performance. While LLaMA is trained on a high-quality and extensive corpus, its deployment in less controlled environments can be more complex due to the need for infrastructure setup, parameter tuning, and task-specific customization.

\vspace{0.07in}

\textit{Use Cases:}
LLaMA embeddings are particularly effective in settings where computational efficiency is critical, such as mobile or edge applications. Their fine-tuning capabilities make them suitable for domain-specific text analysis and multilingual translation tasks, where adapting to specialized vocabularies and nuances is essential \cite{llama_embed_usecase}.

\begin{figure*}[t]
  \centering
  \includegraphics[width=1.5\columnwidth]{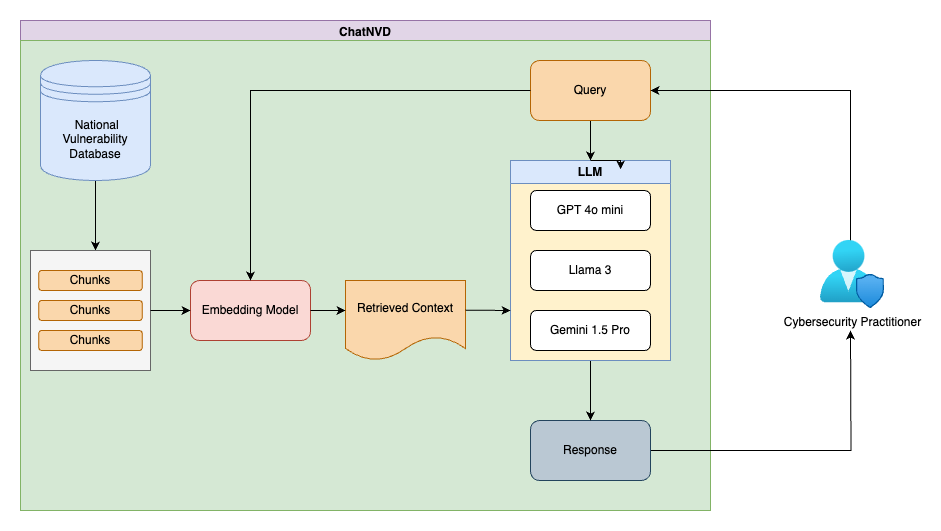}
  \caption{Proposed Architecture of ChatNVD} 
  \label{fig:Architechture_ChatNVD}
\end{figure*}

\vspace{0.07in}

\subsubsection{Gemini Embeddings}
Gemini embeddings are vector representations generated by the Gemini language model developed by Google DeepMind. Designed to encode the semantic meaning of the text into numerical vectors, these embeddings exhibit strong context awareness by dynamically adapting to the surrounding text for more accurate representations. Additionally, the model’s advanced multimodal capabilities enable the integration of text with other data types, such as images or audio, resulting in richer and more comprehensive embeddings \cite{Gemini_multimodal}.

\vspace{0.07in}

\textit{Strengths:}
Gemini embeddings excel at capturing context-dependent meanings, producing nuanced representations that are particularly effective when word interpretation varies with context. Their multimodal integration further enhances their applicability, allowing the model to leverage diverse data sources, a valuable asset in complex application scenarios.

\vspace{0.07in}

\textit{Weaknesses:}
Gemini embeddings demand substantial computational resources for both training and deployment, as they involve high-dimensional multimodal processing and large model sizes \cite{gemini_high_resource}. This can pose significant challenges in resource-constrained environments such as mobile devices, edge computing platforms, or academic research settings with limited GPU access. For instance, deploying Gemini-based models on local servers with modest hardware can lead to slow inference times or memory bottlenecks, making them less practical for real-time or cost-sensitive cybersecurity applications. Moreover, achieving optimal performance may require fine-tuning on specialized, high-quality datasets, which are not always publicly available or easy to curate \cite{high_computation_embedding}. These constraints underscore the importance of considering deployment environments and resource availability when selecting LLM architectures for production use.

\vspace{0.07in}

\textit{Use Cases:}
Gemini embeddings offer strong capabilities for cybersecurity tasks that involve contextual understanding and multimodal data. Their ability to interpret both text and visual inputs makes them effective for summarizing threat reports, detecting phishing attempts, and analyzing vulnerabilities that include linked media such as screenshots or logs. With support for cross-modal reasoning, Gemini can correlate alerts with visual context, improving situational awareness and accelerating incident response \cite{Gemini_multimodal}. These features make Gemini a promising choice for building advanced, multimodal cybersecurity analysis tools.

Based on the comparative insights outlined above, we developed three variants of ChatNVD, each powered by a different LLM: GPT-4o Mini, LLaMA 3, and Gemini 1.5 Pro. These models were selected to reflect diverse strengths: GPT-4o Mini for its cost-efficiency and strong general performance, LLaMA 3 for its open-source adaptability, and Gemini 1.5 Pro for its advanced contextual and multimodal reasoning. To ensure consistency across all variants and reduce computational overhead, we employed a shared TF-IDF embedding pipeline. While modern neural embedding techniques offer richer semantic representations, they are often costly and resource-intensive when applied to large corpora like the NVD. In contrast, TF-IDF provides a lightweight, interpretable, and cost-effective alternative that enables efficient preprocessing without compromising the integrity of comparative analysis. This unified architecture allows for a fair and practical evaluation of each LLM’s effectiveness in the context of real-world cybersecurity vulnerabilities.

\section{Research Methodology} \label{Section 3}

Figure~\ref{fig:Architechture_ChatNVD} presents the system architecture of ChatNVD, designed to assist cybersecurity practitioners in analyzing software vulnerabilities using large language models. The architecture begins with the NVD dataset, which is segmented into smaller, searchable text chunks. These chunks are embedded using a TF-IDF model, selected for its efficiency, transparency, and scalability when processing large corpora. Upon receiving a user query, the system retrieves the most relevant NVD chunks based on embedding similarity. The retrieved context, along with the query, is then passed to a designated large language model, either GPT-4o Mini, LLaMA 3, or Gemini 1.5 Pro, depending on the specific ChatNVD variant in use. The selected model generates a tailored, context-aware response, which is returned to the user to support timely and informed vulnerability assessments.

\begin{figure}[t]
    \adjustbox{scale=1.1}{\includegraphics[width=0.9\columnwidth]{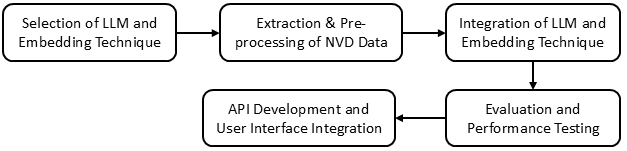}}
    \caption{Phased Workflow of ChatNVD Research Methodology.}
    \label{fig:phase_diagram}
\end{figure}

To operationalize this architecture, we followed a five-phase methodology, as depicted in Figure~\ref{fig:phase_diagram}. The process includes: (1) selecting the LLMs and embedding approach, (2) extracting and preprocessing the NVD data, (3) integrating the embedding model with each LLM, (4) evaluating the performance of the three ChatNVD variants, and (5) developing the API and user interface. This structured approach ensures consistency across variants while supporting scalable, interpretable, and practical deployment in real-world cybersecurity contexts.

\subsection{Phase 1: Selection of LLMs and Embedding Techniques}

In Phase 1, we select the foundational components for ChatNVD: LLMs and the text embedding technique. Based on the comparative analysis presented in Section~\ref{Section 2}, we choose three LLMs, GPT-4o Mini, LLaMA 3, and Gemini 1.5 Pro, each offering distinct advantages relevant to our objectives. GPT-4o Mini is selected for its strong performance–cost trade-off and versatility across general NLP tasks \cite{openai-gpt4o, gpt3.5_good_for_nlp}. LLaMA 3, with its open-source availability and fine-tuning flexibility, supports customizable deployments tailored to cybersecurity contexts. Gemini 1.5 Pro brings advanced multimodal reasoning and strong code understanding capabilities, making it particularly well-suited for vulnerability analysis.

For the embedding layer, we evaluate four techniques, TF-IDF, OpenAI embeddings, LLaMA embeddings, and Gemini embeddings, drawing on the strengths and limitations discussed in Section~\ref{Section 2}. While modern neural embeddings provide rich semantic representations, they introduce significant computational overhead and often require specialized infrastructure. In contrast, TF-IDF offers a lightweight, transparent, and efficient alternative that captures term relevance without requiring large-scale inference or training. Given its low resource requirements and ease of integration, we adopt TF-IDF as the common embedding technique across all ChatNVD variants. This decision enables scalable and responsive processing while preserving sufficient contextual relevance for large-scale vulnerability analysis.

\begin{figure}[t]
\begin{minted}[fontsize=\small, breaklines]{text}

prompt_template = """
        Context: \n{context}.\n
        Task: - You are an assistant that helps with CVE data. \n 
              - Only use the context provided. Respond with CVE details. \n
              - Recommend this website and attach the CVE ID in front of it https://nvd.nist.gov/vuln/detail/ \n
        
        Question: \n{question}?\n

        Answer:\n
        """
\end{minted}
\caption{Prompt template for generating responses in Fig. ~\ref{fig:gemini response}.}
\label{fig:prompt_template_bad}
\end{figure}

\begin{figure}[t!]
  
   \centering
  \includegraphics[width=\columnwidth]{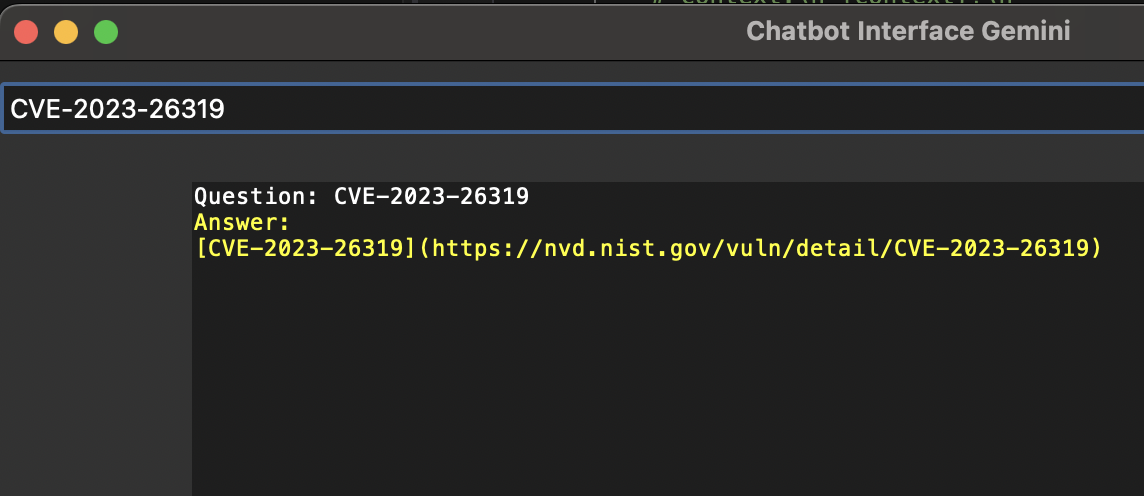}
  \caption{Gemini 1.5 Pro yielding unsatisfactory response  }
  \label{fig:gemini response}
\end{figure}

\subsection{Phase 2: Extraction and Preprocessing of NVD Data}

In Phase 2, we extract and preprocess data from the NVD to create a clean, lightweight knowledge base optimized for use with the TF-IDF embedding model and selected LLMs. The process begins with the removal of duplicate records and non-essential metadata to reduce redundancy and simplify the overall structure. We also reorder JSON keys and values to ensure consistency and facilitate efficient parsing. To further minimize processing overhead, we eliminate formatting noise and trim auxiliary fields that do not contribute to the core vulnerability descriptions. These steps reduce the dataset size from 1.29 GB to 720.7 MB, significantly improving memory efficiency and system responsiveness. By streamlining the dataset while preserving all critical content, we ensure faster context retrieval, reduced inference latency, and overall improved performance across all ChatNVD variants.

\subsection{Phase 3: Integration of LLMs and Embedding Technique}

In Phase 3, we integrate the selected LLMs, GPT-4o Mini, LLaMA 3, and Gemini 1.5 Pro, with the TF-IDF-based retrieval pipeline to establish the core processing loop of ChatNVD. Using the preprocessed and indexed NVD data, the system retrieves the most relevant text segments in response to a user query through TF-IDF similarity scoring. This contextual information is then paired with the original query and fed into the respective language model to generate a concise, informative response. The integration is designed to be modular and consistent across all three model variants, ensuring reliable performance while maintaining system scalability. This phase establishes a clear interface between the retrieval and generation components, enabling ChatNVD to deliver fast, context-aware outputs suitable for real-time vulnerability analysis.


During integration, we also address challenges related to prompt construction, which prove critical to LLM output quality. While we adopt a consistent prompt template to ensure fair evaluation across all models, we observe that prompt effectiveness varies significantly between LLMs. For instance, with Gemini 1.5 Pro, our initial prompt, formed by directly appending the context to the user query (Figure~\ref{fig:prompt_template_bad}), produces incomplete outputs, such as returning only an NVD URL instead of a full CVE description (Figure~\ref{fig:gemini response}). To resolve this, we refine the prompt structure by incorporating clearer task instructions and formatting cues (Figure~\ref{fig:prompt_template_good}). The revised prompt enables Gemini to generate complete, relevant responses that accurately reflect the retrieved context (Figure~\ref{fig:gemini response good}). This experience highlights the importance of prompt tuning and model-specific prompt design, as even minor adjustments can significantly impact the quality, completeness, and reliability of LLM outputs.

\subsection{Phase 4: Evaluation Process}

In Phase 4, we systematically evaluate the performance of the ChatNVD system across its three variants—GPT-4o Mini, LLaMA 3, and Gemini 1.5 Pro—each integrated with TF-IDF embeddings. To ensure a comprehensive and objective assessment, we develop a suite of Python scripts to automate testing using a dataset of 125 cybersecurity-focused questions. As described in Section~\ref{Section 4}, these questions are constructed from five randomized batches of CVEs, with each batch containing five CVEs and five structured questions per CVE, resulting in a total of 125 distinct question–answer pairs. A representative CVE structure used in this process is illustrated in Figure~\ref{fig:cve_json_fixed}.

\begin{figure}[t!]
\begin{minted}[fontsize=\small, breaklines]{text}
prompt_template = """
        You are an assistant that helps with CVE data. Only use the context provided. Respond with CVE details, recommend this website, and attach the CVE ID in front of it https://nvd.nist.gov/vuln/detail/ \n
        
        Context:\n {context}\n
        Question: \n{question}?\n

        Answer:
        """
\end{minted}
\caption{Prompt template for generating responses in Fig. \ref{fig:gemini response good}}
\label{fig:prompt_template_good}
\end{figure}
\begin{figure}[t!] 
   \centering
  \includegraphics[width=\columnwidth]{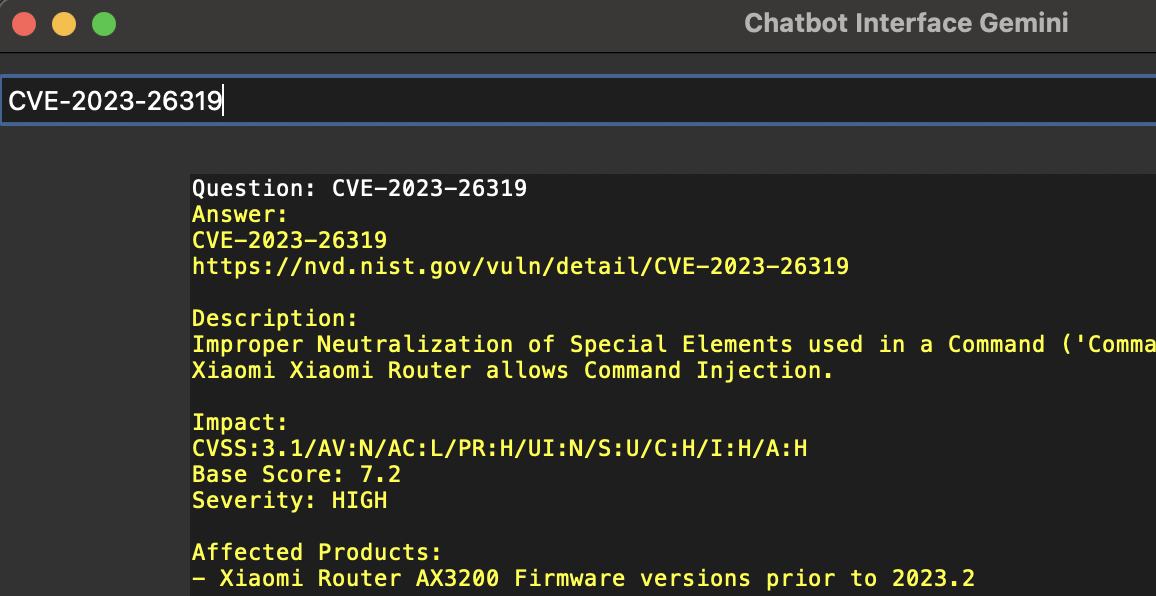}
  \caption{Gemini 1.5 Pro yielding satisfactory response.}
  \label{fig:gemini response good}
\end{figure}

Both the questions and the corresponding ground-truth answers are programmatically derived from the structured fields within the NVD JSON records. For example, a typical question might request the publication date of a given CVE, with the correct answer extracted directly from the \texttt{publishedDate} field of the respective JSON object. This automated approach ensures consistency, eliminates manual bias, and allows for precise comparison between model outputs and verified reference values.

By evaluating the models on real-world vulnerability data, we gain critical insights into their respective strengths and limitations in producing accurate, relevant, and context-aware responses. The results reveal differences in how effectively each LLM interprets structured prompts, extracts key details, and maintains factual alignment with the source data, highlighting important considerations for deploying LLMs in practical cybersecurity applications.

\begin{figure}[H]
\centering
\begin{adjustbox}{max width=\columnwidth}
\begin{minipage}{\linewidth}
\begin{minted}[breaklines, breakanywhere=true, fontsize=\scriptsize]{json}
{
  "cve": {
    "CVE_data_meta": {
      "ID": "CVE-2023-0017",
      "ASSIGNER": "cna@sap.com"
    },
    "problemtype": {
      "problemtype_data": [{
        "description": [{
          "lang": "en",
          "value": "CWE-284"
        }]
      }]
    },
    "description": {
      "description_data": [{
        "lang": "en",
        "value": "An unauthenticated attacker in SAP NetWeaver AS for Java version 7.50, due to improper access control, can attach to an open interface and use a naming and directory API to access services and perform unauthorized operations, including reading, modifying, and disabling services."
      }]
    },
    "references": {
      "reference_data": [{
        "url": "https://launchpad.support.sap.com/#/notes/328093",
        "refsource": "MISC",
        "tags": ["Permissions Required", "Vendor Advisory"]
      }, {
        "url": "https://www.sap.com/documents/2022/02/fa85ea4.html",
        "refsource": "MISC",
        "tags": ["Vendor Advisory"]
      }]
    }
  },
  "configurations": {
    "nodes": [{
      "operator": "OR",
      "cpe_match": [{
        "vulnerable": true,
        "cpe23Uri": "cpe:2.3:a:sap:netweaver_application_server_for_java:7.50:*:*:*:*:*:*:*"
      }]
    }]
  },
  "impact": {
    "baseMetricV3": {
      "cvssV3": {
        "version": "3.1",
        "vectorString": "CVSS:3.1/AV:N/AC:L/PR:N/UI:N/S:U/C:H/I:H/A:H",
        "attackVector": "NETWORK",
        "attackComplexity": "LOW",
        "privilegesRequired": "NONE",
        "userInteraction": "NONE",
        
        "scope": "UNCHANGED",
        "confidentialityImpact": "HIGH",
        "integrityImpact": "HIGH",
        "availabilityImpact": "HIGH",
        "baseScore": 9.8,
        "baseSeverity": "CRITICAL"
      },
      "exploitabilityScore": 3.9,
      "impactScore": 5.9
    }
  },
  "publishedDate": "2023-01-10T04:15Z",
  "lastModifiedDate": "2023-01-13T18:18Z"
}
\end{minted}
\end{minipage}
\end{adjustbox}
\caption{Full CVE entry (CVE-2023-0017) extracted from the National Vulnerability Database.}
\label{fig:cve_json_fixed}
\end{figure}

\subsection{Phase 5: API Development and User Interface Integration}

In Phase 5, we focus on developing a robust API and integrating a user-friendly interface for ChatNVD. Based on the results of the evaluation phase, we select the best-performing LLM and combine it with the TF-IDF embedding pipeline to construct the core ChatNVD API. For backend development, we adopt FastAPI due to its high performance, asynchronous support, and ease of use for building scalable web services \cite{fastapi}. To ensure broad accessibility, we initially deploy the API on an AWS EC2 \texttt{t2.micro} instance under the free tier. However, the limited 1 GB of RAM proves insufficient for processing the LLM queries, leading to frequent crashes. To address this, we upgrade to a \texttt{t2.medium} instance with 4 GB of RAM, which significantly improves stability and ensures reliable API performance under typical usage conditions. The front-end interface is developed using React, a widely adopted JavaScript library known for its flexibility and component-based architecture \cite{React_Intro, React_flexible}. React enables the rapid development of modular, maintainable UI components, allowing us to build a responsive and intuitive interface for end users.

By combining FastAPI, AWS, and React, we deliver a fully functional web application that allows users to submit queries about software vulnerabilities and receive real-time, context-aware responses generated from the NVD. This final phase ensures that ChatNVD is not only technically robust but also accessible and usable by practitioners in real-world cybersecurity environments.

\section{Experimental Results} 
\label{Section 4}

This section presents our comprehensive evaluation of LLMs for vulnerability information retrieval tasks, detailing our experimental setup, dataset characteristics, evaluation methodology, and performance analysis.

\subsection{System Configuration and Experimental Setup}

All experiments were executed on a MacBook Pro equipped with Apple’s M3 Pro chip, featuring an 8-core CPU (4 performance cores at 3.2 GHz and 4 efficiency cores at 2.0 GHz), a 10-core integrated GPU, and 18 GB of unified RAM. This configuration ensured sufficient compute capacity for our evaluation pipeline, including real-time prompt generation, API calls to each LLM, and on-device embedding computations.
\vspace{0.07in}

\subsection{Dataset Description and Preparation}
We utilized the NVD as our primary data source, offering comprehensive and authoritative information on software vulnerabilities. Our dataset encompasses vulnerability records spanning from 2002 to 2024, year-specific JSON files containing chronological vulnerability data, as well as latest revised NVD files to ensure up-to-date vulnerability information.

The data was retrieved on September 18, 2024, resulting in an initial corpus of 1.29GB comprising 2,487 unique vulnerabilities. To optimize the dataset for large language model (LLM) processing, several preprocessing steps were undertaken. These included the elimination of redundant and non-essential fields, the restructuring of complex JSON objects to enhance readability, the removal of unnecessary whitespace and formatting elements, and the standardization of field naming conventions. As a result of these modifications, the dataset size was reduced to 720.7MB, a 44.1\% reduction, while preserving all critical vulnerability information necessary for our evaluation.

\vspace{0.07in}

\textbf{\textit{Step 1: Question Design}}
To comprehensively assess each model's capability in handling vulnerability-related queries, we designed five question types targeting different aspects of vulnerability information:
\begin{enumerate}
\item {Temporal information}: ``What is the published date of CVE-2016-9733?"
\item {Descriptive information}: ``What is the description of CVE-2016-9733?"
\item {Technical metrics}: ``What is the exploitability score of CVE-2016-9733?"
\item {Impact assessment}: ``What is the impact score of CVE-2016-9733?"
\item {Overall severity}: ``What is the base score of CVE-2016-9733?"
\end{enumerate}
These questions were designed to evaluate both factual retrieval capabilities (questions 1, 3, 4, and 5) and contextual understanding (question 2), providing a balanced assessment of LLM performance across different information types.

\textbf{\textit{Step 2: Generation of Testing Batches}}
To ensure robust evaluation across diverse vulnerability profiles, we implemented a stratified random sampling approach. Five CVEs were randomly selected for each batch using a custom Python script. For each selected CVE, a complete data record was extracted from the NVD JSON dataset. The five question types were applied to each selected CVE. Expected answers were automatically extracted from the corresponding NVD records. This process was repeated five times to create five distinct testing batches.
Each batch contained 25 questions (5 questions × 5 CVEs), resulting in a total evaluation corpus of 125 questions across 25 unique vulnerabilities. This approach ensured diversity in vulnerability types, complexity levels, and descriptive content, providing a comprehensive evaluation landscape.

\textbf{\textit{Step 3: LLM Query Process}}
A dedicated Python script was developed to systematically query each LLM with identical inputs. The script:
\begin{itemize}
\item Constructed prompts containing the full vulnerability JSON record followed by a specific question
\item Submitted each prompt to the respective LLM API endpoints
\item Captured and standardized responses for consistent evaluation
\item Generated structured output files containing:
\begin{itemize}
\item The original question
\item The ground truth answer from NVD
\item The verbatim response generated by each LLM
\end{itemize}
\end{itemize}
This process resulted in 15 output files (3 LLMs × 5 batches), each maintaining the same structure to facilitate comparative analysis.

\begin{figure}[t!]
  \centering
  \includegraphics[height=0.27\textwidth]{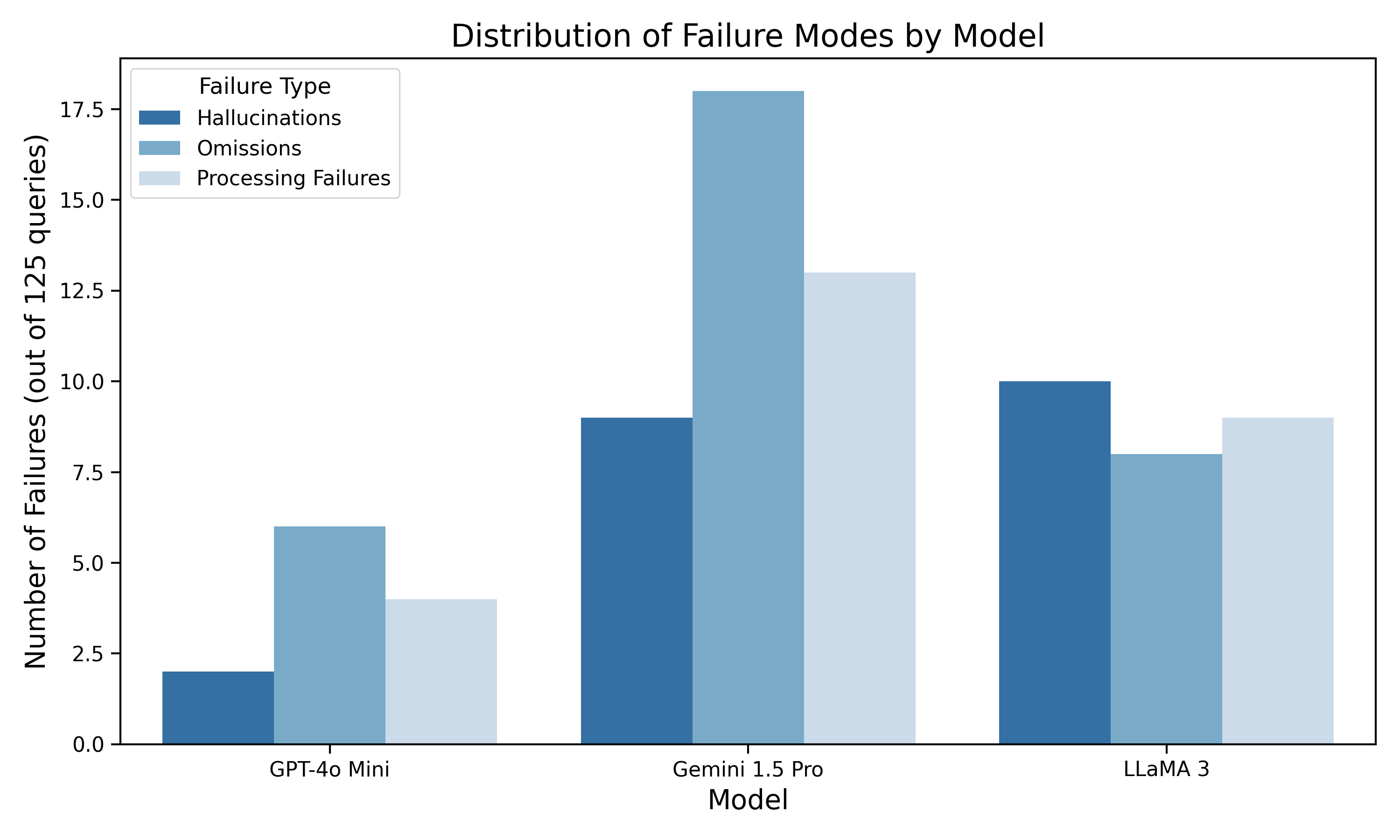}
  \caption{ Distribution of failure modes for all LLMs}
  \label{fig:failure_modes}
\end{figure}

\textbf{\textit{Step 4: Evaluation and Scoring}}
In the final phase, we employed a specialized comparison algorithm to evaluate LLM responses against the ground truth data:
\begin{itemize}
\item \textbf{Accuracy}: Calculated as the proportion of responses that correctly matched the expected answers from the NVD data
\item \textbf{Error Rate}: Measured as the frequency of incorrect or missing responses
\end{itemize}
The error analysis involved categorization of error types into hallucinations (fabricated information), omissions (incomplete information), processing failures (inability to handle specific input types). Figure \ref{fig:failure_modes} presents the distribution of these failure modes across the three evaluated models: GPT-4o Mini, Gemini 1.5 Pro, and LLaMA 3.

GPT-4o Mini exhibited the fewest total failures, with only 2 hallucinations, 6 omissions, and 4 processing failures out of 125 queries. This low error rate highlights the model’s reliability and careful handling of vulnerability-related questions.

Gemini 1.5 Pro, while competitive in overall performance, showed the highest number of total failures. Omissions were the most frequent issue (18 cases), followed by processing failures (13) and hallucinations (9). This suggests that Gemini may struggle with coverage and stability, particularly under complex or ambiguous query conditions.

LLaMA 3 exhibited a different error profile, with hallucinations being the most common (10 instances), followed by 9 processing failures and 8 omissions. This indicates a tendency to generate incorrect or misleading content more often than the other models while also facing challenges with processing robustness. The analysis shows GPT-4o Mini as the most reliable model across all failure types, while Gemini 1.5 Pro and LLaMA 3 reveal differing but significant weaknesses, Gemini with omissions and LLaMA with hallucinations, emphasizing the importance of aligning model choice with specific task requirements and risk tolerance.

The performance metrics were compiled into comprehensive reports for each model, enabling detailed comparative analysis.

\begin{figure}[t!]
  \centering
  \includegraphics[height=0.28\textwidth]{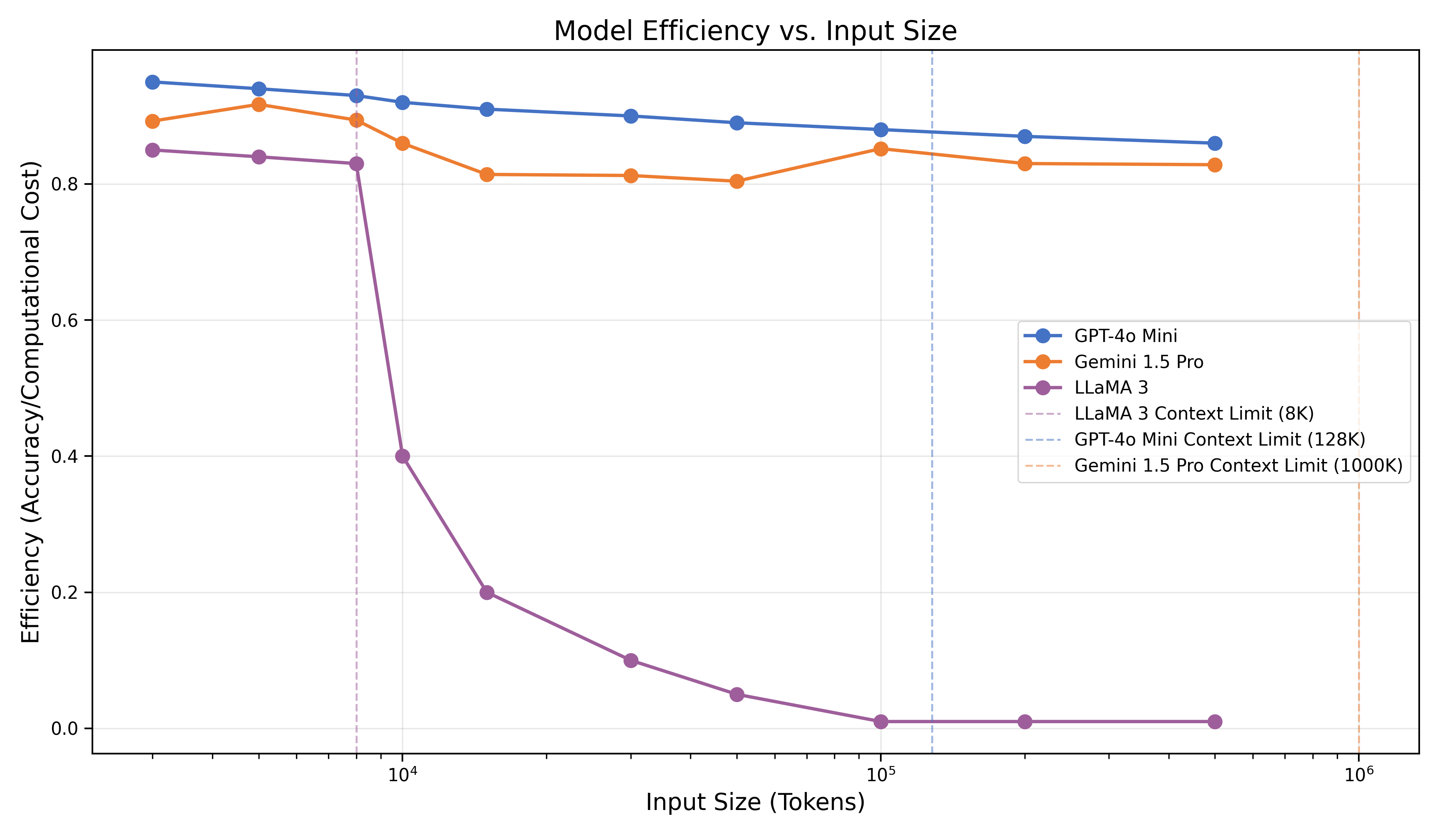}
  \caption{Efficiency of each language model as a function of input size (token count).}
  \label{fig:efficiency_by_input_size}
\end{figure}

\subsection{Summary of Results}
We evaluated the performance of three LLMs, GPT-4o Mini, LLaMA 3, and Gemini 1.5 Pro, on a total of 125 security vulnerability questions. These questions were distributed across five batches, with each batch comprising 25 questions targeting five unique and randomly selected CVEs. The combined accuracy and error rates for all batches are presented in Table \ref{tabb}. The performance of GPT-4o Mini, Gemini 1.5 Pro, and LLaMA 3 across five evaluation batches is summarized. GPT-4o Mini achieved superior accuracy in all batches. Gemini 1.5 Pro exhibited performance fluctuations, with a significant drop observed in Batch 5. LLaMA 3 also performed well overall but exhibited variability, particularly in the third batch.
\begin{table}[t]
\centering
\caption{Accuracy of LLMs Across Batches of CVE-Based Queries.}
\begin{tabular}{lccccc}
\toprule
\textbf{Model} & \textbf{Batch 1} & \textbf{Batch 2} & \textbf{Batch 3} & \textbf{Batch 4} & \textbf{Batch 5} \\
\midrule
GPT-4o Mini     & 0.97 & 0.98 & 0.92 & 0.94 & 0.96 \\
Gemini 1.5 Pro  & 0.92 & 0.78 & 0.84 & 0.88 & 0.68 \\
LLaMA 3         & 0.83 & 0.80 & 0.52 & 0.90 & 0.70 \\
\bottomrule
\end{tabular}
\label{tabb}
\end{table}

GPT-4o Mini achieved consistently high accuracy across all five batches, with scores tightly clustered between 0.92 and 0.98, indicating strong and stable performance. Gemini 1.5 Pro performed well in Batch 1 (0.92) but exhibited noticeable fluctuations thereafter, with a significant dip in Batch 5 (0.68), suggesting potential inconsistencies in handling certain question types. LLaMA 3 showed a more erratic pattern, with relatively strong results in Batches 1, 2, and 4, but a sharp drop in Batch 3 (0.52). Despite recovering somewhat in Batch 5, this variability points to less consistent performance compared to the other models.

\subsubsection{Performance Analysis by Input Complexity}
Figure \ref{fig:efficiency_by_input_size} illustrates how model efficiency, defined as the ratio of accuracy to computational cost, varies with increasing input size (token count) for GPT-4o Mini, Gemini 1.5 Pro, and LLaMA 3. GPT-4o Mini demonstrates the most consistent efficiency across input sizes, showing only a slight downward trend as input complexity increases. Its ability to maintain relatively high efficiency up to its 128K token context limit suggests strong scalability and robustness for tasks involving large contexts. Gemini 1.5 Pro also maintains stable performance across a wide input range, albeit with slightly lower efficiency than GPT-4o Mini. It exhibits a small dip around mid-range token sizes but recovers slightly at higher inputs, operating effectively even as it approaches its 1M token context limit.

In contrast, LLaMA 3 experiences a steep drop in efficiency shortly after surpassing its 8K token context limit. Its performance rapidly deteriorates as the input size increases, indicating significant challenges in handling longer contexts. Beyond 10K tokens, its efficiency becomes negligible, highlighting limitations in both scalability and computational cost-effectiveness for complex inputs.
GPT-4o Mini proves to be the most efficient and stable across a wide range of input sizes, making it well-suited for high-complexity tasks. Gemini 1.5 Pro is moderately effective but shows more variability, while LLaMA 3 struggles considerably beyond its relatively limited context window.

\subsubsection{Performance Analysis by Question Type}
Analysis of performance by question type revealed distinct model behaviours:
\begin{figure}[t]
  \centering
  \includegraphics[height=0.27\textwidth]{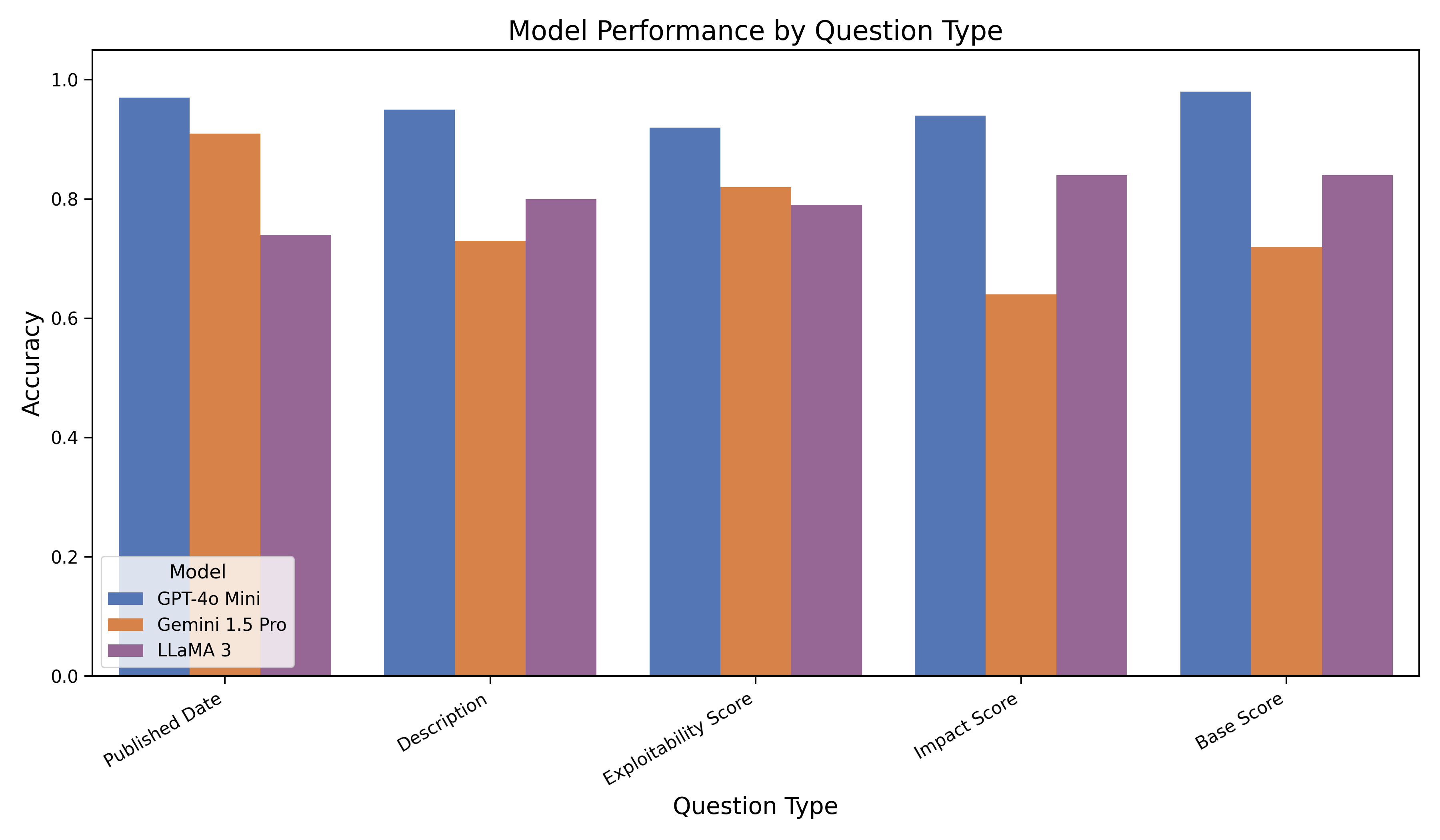}
  \caption{Performance on different types of questions}
  \label{fig:question_type_performance}
\end{figure}
\begin{itemize}
\item \textbf{Descriptive Questions}: GPT-4o Mini provided complete and accurate responses to all 25 descriptive questions. In contrast, Gemini 1.5 Pro returned blank outputs for 18 out of 25 descriptive questions despite generating responses for other question types. LLaMA 3 provided partial or incomplete responses to descriptive queries.
\item \textbf{Numerical Metric Questions} (exploitability, impact, and base scores): GPT-4o Mini maintained superior accuracy. LLaMA 3 exhibited hallucination issues, providing incorrect numerical values in 2 cases and failing to generate meaningful responses in 4 instances. Gemini 1.5 Pro showed inconsistent performance with numerical metrics.
\end{itemize}
Figure \ref{fig:question_type_performance} illustrates how each model performs across various vulnerability-related question categories: Published Date, Description, Exploitability Score, Impact Score, and Base Score.
Across all categories, GPT-4o Mini consistently outperforms both Gemini 1.5 Pro and LLaMA 3, demonstrating high and stable accuracy ranging from 0.93 to 0.98. Its top performance in Base Score (0.98) and Published Date (0.97) questions suggests strong fact recall and structured reasoning capabilities, which are critical for vulnerability assessments.
Gemini 1.5 Pro shows mixed performance. It performs reasonably well on Published Date and Exploitability Score questions but underperforms significantly in the Impact Score category, where its accuracy drops to 0.64. This inconsistency may reflect difficulties in interpreting contextual severity or handling nuanced score calculations.

LLaMA 3 displays relatively uniform performance across most question types, with scores in the 0.74–0.85 range. It achieves its best accuracy on Impact Score and Base Score questions (both 0.84), which is notable given its lower scores in earlier evaluations. However, it underperforms compared to GPT-4o Mini and shows more variability than Gemini 1.5 Pro in certain categories.
These results highlight GPT-4o Mini’s superior generalization across question types while suggesting Gemini’s weaknesses in analytical scoring tasks and LLaMA 3’s moderate but steadier, performance across the board. These findings reinforce the need to align model selection with the specific demands of question type in real-world vulnerability analysis contexts.

\begin{figure}[t]
  \centering
  \includegraphics[height=0.22\textwidth]{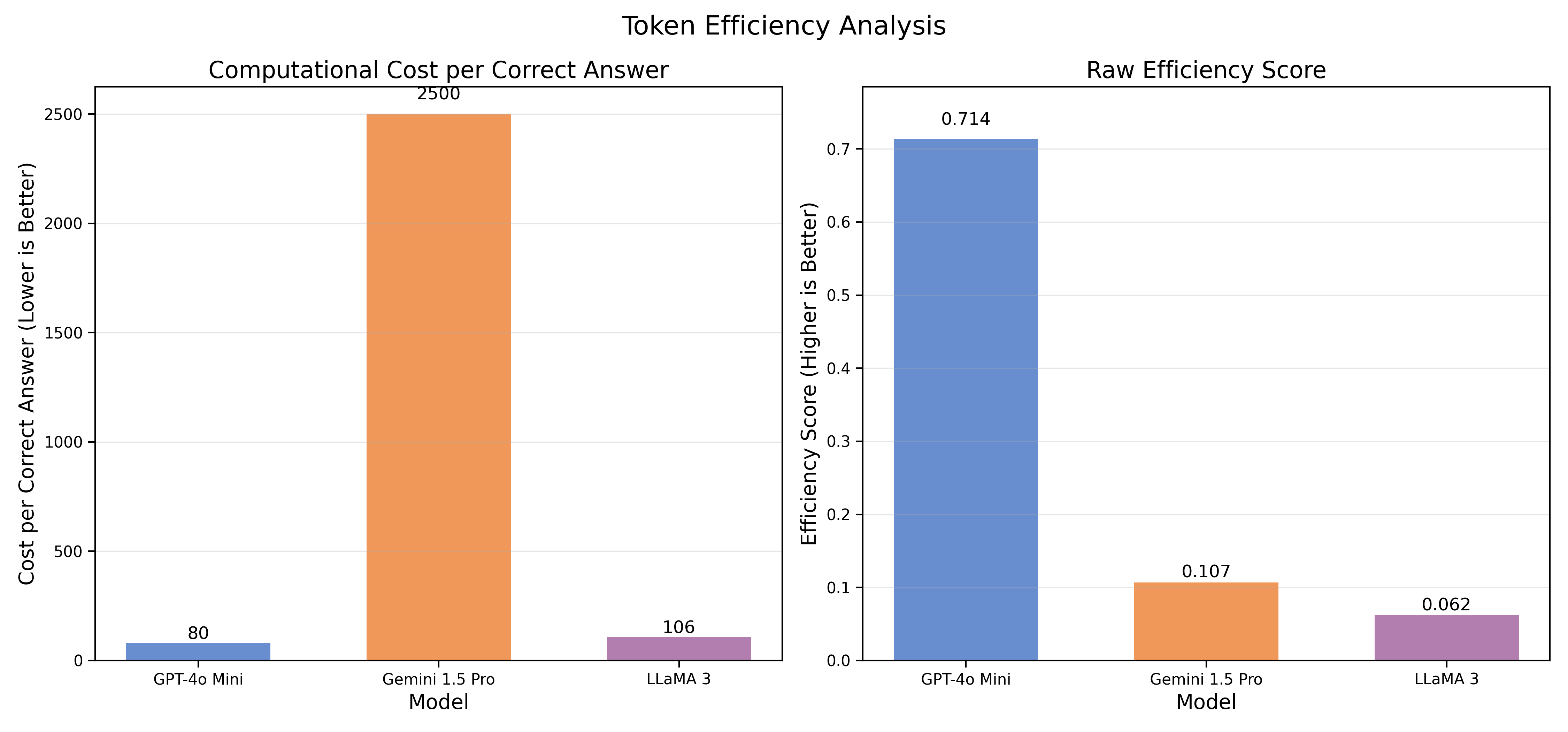}
  \caption{Analyzing token efficiency}
  \label{fig:token_efficiency_analysis}
\end{figure}

Figure \ref{fig:token_efficiency_analysis} clearly demonstrates significant disparities in token efficiency across three major language models: GPT-4o Mini, Gemini 1.5 Pro, and LLaMA 3. The analysis presents two critical metrics: Computational Cost per Correct Answer and Raw Efficiency Score.
In terms of computational cost, GPT-4o Mini demonstrates remarkable efficiency with only 80 units per correct answer, while Gemini 1.5 Pro requires approximately 2500 units i.e. over 31 times higher. LLaMA 3 shows moderate efficiency at 106 units, slightly higher than GPT-4o Mini but substantially more efficient than Gemini 1.5 Pro.
The Raw Efficiency Score further emphasizes GPT-4o Mini's superior performance (0.714), which is approximately 6.7 times higher than Gemini 1.5 Pro (0.107) and 11.5 times higher than LLaMA 3 (0.062). This suggests that GPT-4o Mini achieves significantly better resource utilization while maintaining accuracy.
These findings indicate that architectural design choices in smaller models like GPT-4o Mini may offer valuable insights for developing more resource-efficient AI systems without compromising performance, a critical consideration for real-world deployment scenarios where computational resources may be constrained.

\subsubsection{Error Analysis}
Our error analysis revealed specific limitations in model capabilities:
\begin{itemize}
\item \textbf{Context Window Limitations}: LLaMA 3's 8K token context window proved insufficient for processing larger vulnerability descriptions, resulting in incomplete or incorrect responses.
\item \textbf{Structured Data Processing}: Both LLaMA 3 and Gemini 1.5 Pro demonstrated difficulties in accurately extracting and interpreting structured data from complex JSON objects.

\item \textbf{Descriptive Query Handling}: Gemini 1.5 Pro exhibited a consistent pattern of failure with descriptive questions, suggesting fundamental limitations in its ability to process and synthesize descriptive information from context.

\item \textbf{Numerical Accuracy}: LLaMA 3 showed vulnerability to hallucinations when processing numerical metrics, raising concerns about its reliability for security-critical applications.
\end{itemize}
These findings demonstrate that beyond raw context window size, factors such as model architecture, token utilization strategies, and training objectives significantly impact an LLM's ability to process and interpret vulnerability information effectively.

\subsection{Discussion}

Our evaluation of LLMs revealed significant performance variations when processing vulnerability-related queries, particularly in handling descriptive content and large-context inputs. The following analysis provides insights into the models' capabilities and limitations.

\subsubsection{Performance Across Different Input Sizes}

A critical challenge emerged with \textit{CVE-2022-33248}, which contained a large JSON object of approximately 31,415 tokens. Both LLaMA 3 and Gemini 1.5 Pro failed to answer any questions related to this CVE, including the published date, description, and various security scores (exploitability, impact, and base scores).

LLaMA 3's failure is attributable to its 8K token context window \cite{llama3-context-window}, which is insufficient for processing such large inputs. Surprisingly, Gemini 1.5 Pro also failed despite having a substantially larger context window comparable to GPT-4o Mini. In contrast, GPT-4o Mini with its 128K token context window \cite{gpt4o-mini-context-window} processed the same input successfully, demonstrating that effective large-context processing depends not only on maximum context size but also on model architecture, token utilization strategies, and training objectives.

This pattern persisted even with smaller inputs. Both LLaMA 3 and Gemini 1.5 Pro struggled with \textit{CVE-2016-9733}, which contained only 3,235 tokens—well within their theoretical context windows. While LLaMA 3 could identify relevant contextual keywords (e.g., "IBM"), it still failed to provide accurate responses, suggesting limitations in structured data processing rather than context window constraints.

\subsubsection{Challenges with Descriptive Queries}

Gemini 1.5 Pro exhibited a consistent weakness in handling descriptive questions. Of the 25 descriptive questions across all batches, 18 resulted in blank outputs despite the model generating detailed responses for other query types. This performance did not improve with question rephrasing, indicating fundamental limitations in how the model processes descriptive prompts or leverages contextual information. 

In contrast, GPT-4o Mini provided consistent and complete responses to all descriptive questions. This disparity highlights the importance of effective contextual encoding strategies for descriptive information processing, suggesting that further fine-tuning and improved training data for descriptive tasks could address these limitations in models like Gemini 1.5 Pro.

\subsubsection{Hallucinations and Numerical Precision}

LLaMA 3 demonstrated hallucination issues, specifically when responding to impact score questions. From 25 queries, it provided incorrect numerical values in two cases and failed to generate meaningful responses in four instances. These hallucinations raise concerns about the model's numerical reasoning and factual accuracy when processing structured cybersecurity data, suggesting the need for robust validation mechanisms to mitigate such errors.

\begin{figure*}[t]
  \centering
  \includegraphics[width=1.5\columnwidth]{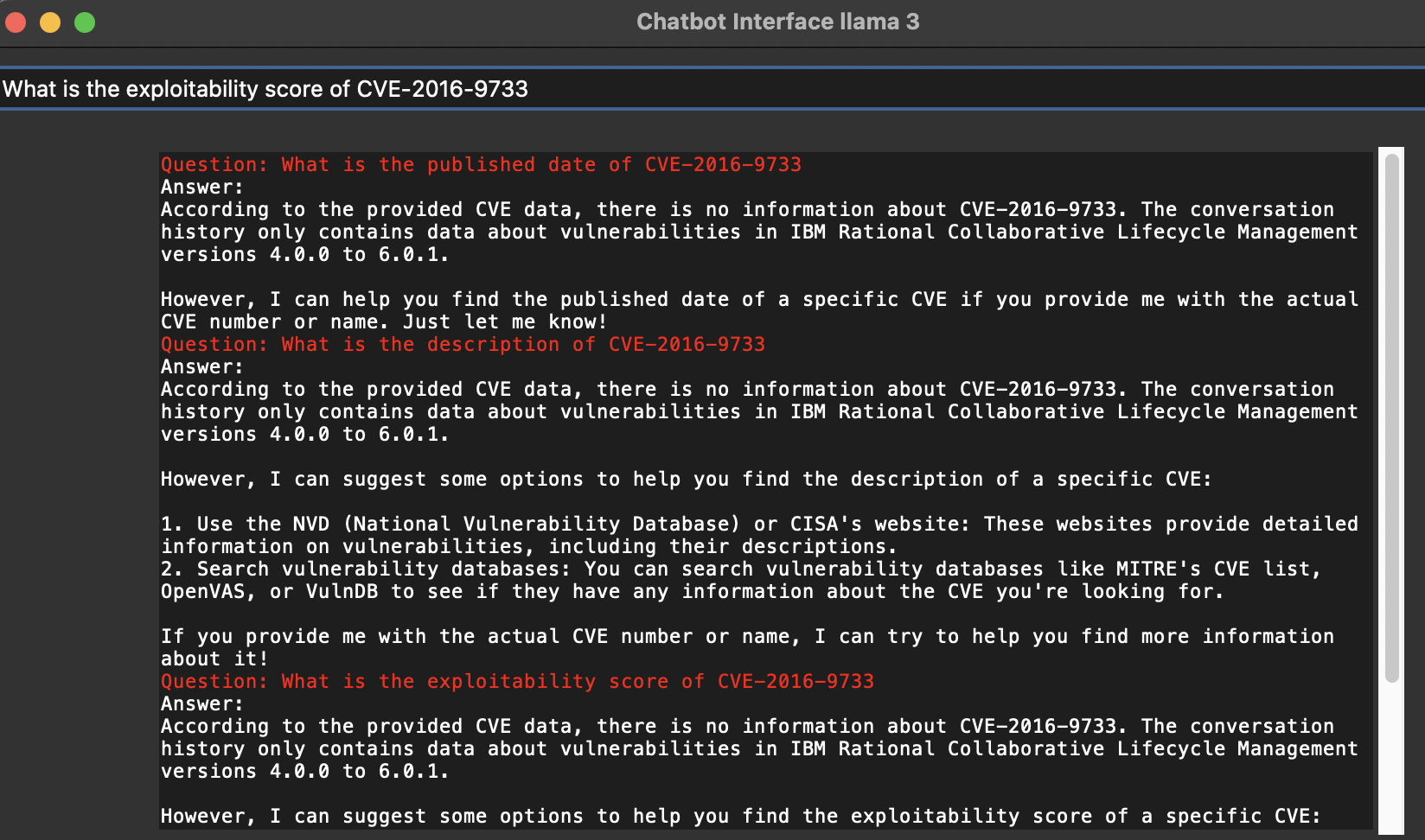}
  \caption{Case study highlighting LLaMA 3’s failure to correctly respond to queries related to CVE-2016-9733, despite contextual availability.} 
  \label{fig:llama response}
\end{figure*}

\begin{figure*}[t]
  \centering
  \includegraphics[width=1.5\columnwidth]{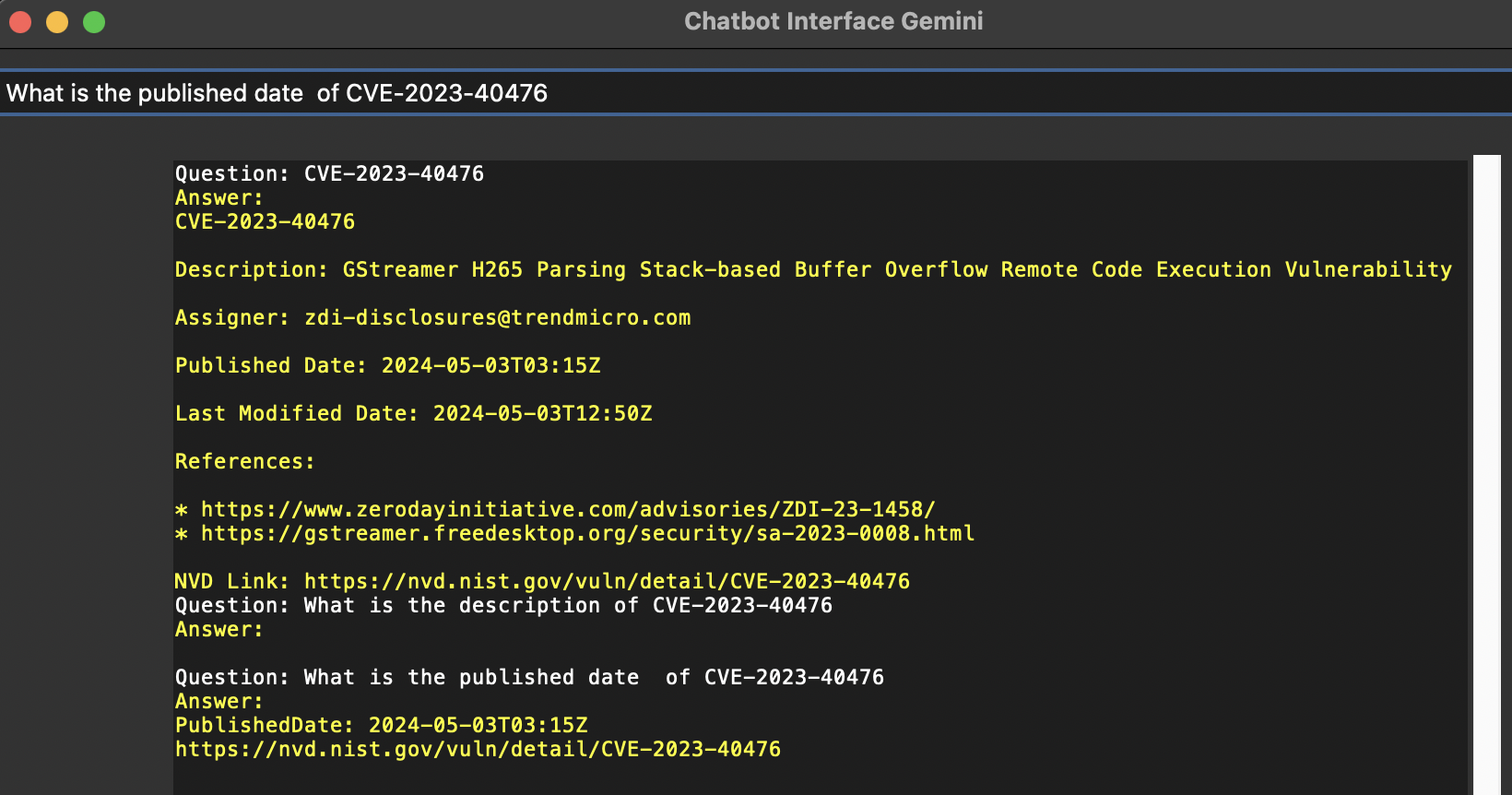}
  \caption{Illustration of Gemini 1.5 Pro’s failure to return a meaningful description for CVE-2023-4047, despite extracting other metadata accurately.} 
  \label{fig:Gemini blank response}
\end{figure*}

\subsubsection{Key Takeaways}

Our evaluation yields several valuable insights for researchers and practitioners:

\begin{itemize}
    \item \textbf{Reliability and Consistency:}  
    GPT-4o Mini's superior accuracy across all batches establishes it as highly reliable for cybersecurity tasks requiring consistent information retrieval. Its ability to handle diverse question types makes it particularly suitable for comprehensive vulnerability assessments.

    \item \textbf{Evaluation Methodology:}  
    The variable performance of LLaMA 3 and Gemini 1.5 Pro underscores the importance of evaluating LLMs with domain-specific queries of varying complexity. Comprehensive testing that includes large-context and descriptive prompts ensures models can handle real-world cybersecurity scenarios involving complex structured data.

    \item \textbf{Hallucination Mitigation:}  
    The hallucinations observed in LLaMA 3 responses highlight the need for robust model validation. Implementing confidence-based outputs, external reference cross-checking, and fine-tuned training datasets for numerical data could improve factual accuracy in structured queries.

    \item \textbf{Resource Optimization:}  
    While models with large context windows like GPT-4o Mini offer superior performance, they incur significant computational costs. Organizations should consider alternatives such as smaller fine-tuned models or efficient context-chunking strategies to optimize resource usage without compromising essential capabilities.

    \item \textbf{Task-Specific Metrics:}  
    The varied performance across query types indicates that evaluation frameworks should incorporate diverse query categories, metadata retrieval, numerical scores, and descriptive content to comprehensively assess LLM strengths and weaknesses. Cybersecurity-specific performance metrics can further guide model selection and development.
\end{itemize}

Figure \ref{fig:llama response} LLaMA 3 demonstrates notable limitations when queried about CVE-2016-9733. Despite multiple reformulations of the question (regarding published date, description, and exploitability score), LLaMA 3 consistently responds with the same pattern: claiming no information exists while referencing unrelated IBM Rational Collaborative Lifecycle Management vulnerabilities. This suggests a potential knowledge gap or retrieval issue within LLaMA 3's training data specifically related to this CVE.
Similarly, while Gemini 1.5 Pro provides detailed information about CVE-2023-4047 Figure \ref{fig:Gemini blank response} (including description, assigner, publication date, and references), it appears to deliver an incomplete response when asked specifically about the CVE description.
\section{Implications for Researchers and Practitioners}\label{implication}

Our findings offer significant implications for both research and applied contexts:
\begin{enumerate}
    \item \textit{Model Selection for High-Stakes Applications:}
    In critical applications such as vulnerability assessment and threat detection, reliability should be prioritized over other performance metrics. The consistent accuracy demonstrated by GPT-4o Mini makes it suitable for scenarios requiring dependable information retrieval from complex vulnerability data.

    \item \textit{Domain-Specific Testing:}
    General-purpose evaluation is insufficient for assessing LLM performance in specialized domains like cybersecurity. Testing should incorporate domain-specific queries that reflect real-world use cases, including vulnerability descriptions, exploitability scores, and impact metrics, to identify model limitations in processing complex contextual data.

    \item \textit{Custom Evaluation Frameworks:}
    Our methodology suggests that other cybersecurity applications, such as threat intelligence and malware analysis, would benefit from tailored evaluation approaches. Researchers should develop task-specific metrics that assess an LLM's ability to process technical and structured data accurately, ensuring alignment with specific cybersecurity requirements.

    \item \textit{Performance-Resource Balance:}
    Practitioners must consider trade-offs between performance and computational demands, particularly with large-context models like GPT-4o Mini. In resource-constrained environments, optimized smaller models may offer a more cost-effective alternative if properly fine-tuned for specific tasks.

    \item \textit{Addressing Failure Patterns:}
    The consistent failure patterns observed, such as Gemini 1.5 Pro's blank responses to descriptive questions, indicate that model architecture and token management strategies significantly influence performance beyond context size. Researchers should analyze these failure modes to enhance model robustness through improved training data, refined token prioritization, and post-processing strategies.
\end{enumerate}

\begin{figure}[t]
  \centering
  \includegraphics[width=\columnwidth, scale=0.8]{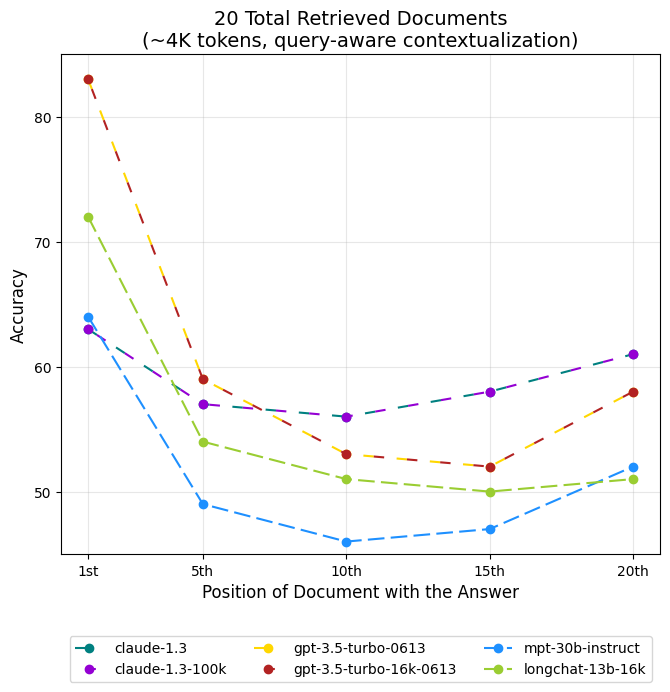}
  \caption{Performance on Lost-in-the-middle problem \cite{liu2024lost}.}
  \label{fig:loss in the middle problem}
\end{figure}

\section{Threats to Validity} \label{Section 5}

While ChatNVD demonstrates promising performance, several factors threaten the validity of its outputs. The following sections detail these challenges.

\begin{enumerate}[leftmargin=*]

\item{\textit{Response Hallucination:}} One concern is response hallucination, wherein the LLM generates outputs that extend beyond or deviate from the provided context and established knowledge base \cite{ji2023survey}. Such erroneous or fabricated responses can compromise the reliability of ChatNVD and erode user trust. Mitigation strategies include enhancing training datasets, performing targeted fine-tuning to improve contextual grounding, and incorporating robust post-processing validation mechanisms that cross-reference generated content with trusted data sources. \vspace{0.075in}

\item{\textit{Contextual Token Limitations:}} ChatNVD relies on maintaining contextual continuity throughout interactions to produce accurate and pertinent responses. However, the inherent token limitations in current LLM architectures may cause the model to lose track of earlier context as the conversation length increases \cite{liu2024lost}. This loss of context can lead to inconsistencies or a decline in response relevance. Possible solutions include optimizing token allocation, employing context summarization techniques, and adjusting the model’s attention mechanisms to better capture long-range dependencies. \vspace{0.075in}

\item{\textit{Lost-in-the-Middle Phenomenon:}} During our experiments with long vulnerability entries, we observed a recurring pattern: model accuracy remained relatively high (70–80\%) for information located at the beginning or end of an input sequence, but dropped sharply (0–50\%) for content appearing in the middle (see Figure~\ref{fig:loss in the middle problem}). This pattern aligns with what recent studies refer to as the “lost-in-the-middle” phenomenon \cite{liu2024lost}, wherein transformer-based models struggle to retain and prioritize information positioned in the center of long inputs. While not a core focus of this paper, this behaviour emerged frequently in our evaluations and highlights the need for improved prompt structuring or context window utilization strategies in future systems. \vspace{0.075in}

\item{\textit{Operational Costs and Scalability:}} The high operational costs associated with token-based pricing models pose another significant threat to the long-term viability of ChatNVD. Each query incurs a cost proportional to the number of tokens processed, and as the system scales, these costs can become prohibitive. While transitioning to a subscription-based or paid model may offset some expenses, it could also restrict accessibility and user adoption. Therefore, exploring cost-effective model optimizations and sustainable funding strategies is essential for ensuring that ChatNVD remains both economically viable and widely accessible.
\end{enumerate}

\section{Conclusion}
\label{sec:conclusion}
We proposed ChatNVD, an LLM-based cybersecurity vulnerability assessment tool built on the NVD dataset to deliver an interactive and efficient solution for vulnerability analysis. ChatNVD provided detailed, context-rich insights that streamlined the assessment process for diverse users, including cybersecurity professionals, developers, and non-technical stakeholders. To achieve this, we developed three variants of ChatNVD, using GPT-4o Mini, Gemini 1.5 Pro, and LLaMA 3, integrated with the TF-IDF embedding technique and leveraging NVD JSON feeds (2002–2024) as the foundational knowledge. We also developed a tailored evaluation framework to measure each model’s accuracy and error rates. Our evaluation across 125 CVE-based queries showed that GPT-4o Mini consistently outperformed Gemini 1.5 Pro and LLaMA 3, achieving accuracy scores of over 92\% across all batches, demonstrating high reliability and computational efficiency. It also exhibited the fewest hallucinations, strongest descriptive reasoning, and most consistent performance across diverse input sizes and question types. Based on these results, we selected GPT-4o Mini for our web application. We then implemented the system using accessible APIs and a user-friendly React interface, and successfully deployed it for real-world use.

For future work, exploring various embedding models that capture deeper semantic meaning in text remains a priority. Incorporating and evaluating additional LLMs, particularly emerging models with larger context windows or domain-specific training, could provide insights into their potential to outperform GPT-4o Mini. Expanding the application’s use cases within the cybersecurity domain could also yield promising results, broadening its applicability and impact. 



\vspace{0.05in}

\bibliographystyle{ieeetr}
\bibliography{ref}

\end{document}